\documentclass[twocolumn,prb]{revtex4}
\usepackage{graphicx}
\usepackage{dcolumn}
\usepackage{bm}
\usepackage{amsmath}

\begin{document}

\title{Magneto-optical response of CdSe nanostructures}
\author{Po-Chung Chen and K. Birgitta Whaley} \affiliation{Department
of Chemistry and Pitzer Center for Theoretical Chemistry, University
of California Berkeley}

\date{\today}

\begin{abstract}
We present theoretical calculations of the Land\'e g-factors of
semiconductor nanostructures using a time-dependent empirical
tight-binding method. The eigenenergies and eigenfunctions of the band
edge states are calculated as a function of an external magnetic field
with the electromagnetic field incorporated into the tight-binding
Hamiltonian in a gauge-invariant form. The spin-orbit interaction and
magnetic field are treated non-perturbatively. The g-factors are
extracted from the energy splitting of the eigenstates induced by the
applied magnetic field.  Both electron and hole g-factors are
investigated for CdSe nanostructures. The size and aspect ratio
dependence of g-factors is studied. We observe that the electron
g-factors are anisotropic and find that the calculated values agree
quantitatively with experimental data. We conclude that the two
distinct g-factor values extracted from time resolved Faraday rotation
experiments should be assigned to the anisotropic in plane
($g_\parallel$) and out of plane ($g_z$) electron g-factors, rather
than to isotropic electron and exciton g-factors.  We find that the
anisotropy in the electron g-factor depends on the aspect ratio of the
nanocrystal. The g-factor anisotropy derived from the wurtzite
structure and from the non-unity aspect ratio may cancel each other in
some regime. We observe that hole g-factors oscillate as a function of
size, due to size dependent mixing between the heavy hole-light hole
components of the valence band edge states. Extension to the
calculation of exciton g-factors is discussed.
\end{abstract}
\maketitle

\section{introduction}
Spin dynamics in semiconductor nanostructures have been studied
intensively in recent years, motivated by the emerging field of
semiconductor spintronics and quantum information
processing.\cite{young02} The most important time scale when
implementing the quantum computer is the decoherence time of the
quantum degree of freedom which is intended to be used as the qubit.
Typically the spin decoherence time in the bulk semiconductor material
is extremely short. However, it is expected that the spin decoherence
time should increase substantially in nanostructures due to the
three-dimensional quantum confinement. This expectation is supported
by the optical orientation experiments of Gupta {\it et al.}
\cite{gupta02} where a nano-second spin lifetime was measured for
neutral CdSe nanostructures. This indicates that there will be plenty
of time to perform quantum operations on the spin degree of freedom in
semiconductor nano structure before the coherence is
lost. Consequently spins in nanostructures are excellent candidates
for qubits.  On the other hand, both spin based quantum computation
and spintronics require precise control of the spin. Since the control
of the spin dynamics in nanostructures is strongly dependent on the
g-factors of electrons, holes, and excitons in the nanostructure, it
is imperative to understand the behavior and magnitude of g-factors.

Experimentally the g-factors of CdSe nanocrystals with wurtzite
lattice structure have been measured via Time Resolved Faraday
Rotation (TRFR) \cite{gupta99,gupta02} and Magnetic Circular Dichroism
(MCD). \cite{kuno98} The TRFR experiment measurement reveals multiple
g-factors. Two or four distinct g-factor values are extracted,
depending on the size of the nanostructure which ranges from 22 \AA\
to 80 \AA\ in diameter. The size distribution of the sample is about
5-15\%.\cite{kadavanich97,gupta99} The sample from 22 \AA\ to 57 \AA\
has a size-dependent mean aspect ratio ranging from 1.0 to 1.3, with a
$\pm 0.2$ variation \cite{kadavanich97,gupta99,joshua03}.  It was
suggested that when there are four distinct g-factors, these g-factors
should be assigned to anisotropic electron and exciton
g-factors. \cite{gupta02} It has been speculated that there exists a
quasispherical regime in which the g-factors become isotropic, hence
reducing the number of distinct g-factors from four to two. On the
other hand, MCD measurements reveal only a single exciton g-factor for
nanocrystals with 19 \AA \ and 25 \AA \ in diameter.\cite{kuno98} It
should be noted that the hole spin is also initially aligned by the
optical pumping in a TRFR experiment. It has been argued that the fast
decoherence of the hole spin makes it impossible to detect the hole
g-factor in TRFR.\cite{gupta02} Although it is well established
experimentally \cite{hilton02} and theoretically \cite{uenoyama90}
that the hole spin decoherence time in the bulk semiconductor is
extremely small, the three-dimensional quantum confinements might also
alter significantly the hole spin decoherence time in the
nanostructure.  To the best of our knowledge there is neither
experimental measurement nor theoretical estimation about the hole
decoherence time in those nanostructures.  Recent time-resolved
photoluminescence on InAs/GaAs quantum dots \cite{paillard01} suggests
that neither the electron nor the hole spin relax on the lifetime
scale of the exciton in the system but no estimation of the hole
relaxation time could be made. This suggests that a hole decoherence
time becomes much longer in nanostructures.  It is thus not yet clear
whether the hole g-factor signature should appear in a TRFR
experiment.

Theoretically, the size dependence of the electron g-factor in CdSe
nanostructures has been calculated within the eight band Kane model
\cite{kiselev98,rodina02_1,rodina02_2}, and in the tight-binding
model.  \cite{joshua03, nenashev03} In general, the effective mass
approximation (EMA) type calculation is inadequate for nanostructures
at small sizes ( $\leq$ 30\AA\ ) because the atomic nature and surface
effects become more prominent as the size of the nanostructure
decreases. Because of its atomic nature, the tight-binding model is
ideal to study the electronic and optical properties of nanostructures
in this size range.

The time-independent tight-binding approach to calculation of
g-factors in Ref 14 is based on Stone's formula \cite{stone63} which
is derived form the double second order perturbation in terms of the
spin-orbit interaction and the external magnetic field. The results
obtained from this perturbative analysis show strong shape
dependence. It was observed that a transition from anisotropic to
isotropic g-factor tensor occurs at aspect ratio $\approx 0.3$,
resulting in a quasispherical regime, as originally suggested by
Rodina and coworkers. \cite {rodina02_2} Since the spin-orbit
interaction is strong in CdSe, ($\lambda_{\mbox{Cd}}=0.151$eV and
$\lambda_{\mbox{Se}}=0.320$eV), it is desirable to include this
non-perturbatively in order to get quantitative values. This not only
provides a more accurate estimation of the electron g-factors for
nanostructures with strong spin-orbit interaction, but will also
enable us to treat the hole and exciton g-factors systematically, in
addition to the electron g-factors. It is also intriguing to
investigate the possibility of a quasispherical regime in more a
quantitative calculations for the electron g-factor.

In this paper we present such non-perturbative theoretical
calculations of the electron and hole g-factors for CdSe
nanostructures employing the time-dependent empirical tight-binding
theory. Both the spin-orbit interaction and the external magnetic
field are taken into account non-perturbatively. The g-factors are
extracted from the magnetic field induced energy shifts of the
electron and hole eigenvalues. The size and aspect ratio dependence of
the resulting g-factors is investigated. We observe that the electron
g-factors decrease monotonically as a function of size and are
strongly anisotropic.  The calculated values agree well with the
experimental data from TRFR. It will be shown that the electron
g-factors can explain the TRFR experimental data without invoking a
exciton g-factor. The result also shows partial cancellation between
the anisotropy deriving from the wurtzite structure and from the
aspect ratio. The hole g-factors show very different behavior. They
show marked oscillations as a function of the size. This is due to the
size sensitive mixing between the heavy hole and light hole
components. The exciton g-factor is not included in the current
calculation. However the same calculation scheme can be easily
extended to calculate the exciton g-factor with Coulomb interactions
included non-perturbatively. \cite{hill96, leung97}

The paper is organized as follows: In section \ref{theory} we
summarize the empirical tight-binding Hamiltonian for CdSe
nanostructures and describe how to solve the model using a
time-dependent approach. In section \ref{results} we present our
numerical results, including total density of states, band gap, and
electron and hole g-factors. In section \ref{sum} we draw several
conclusions and discuss possible future directions.

\section{Theory and Model}
\label{theory}
%In this section we will first construct the tight-binding model of
%CdSe nanostructure. We then describe briefly the time-dependent
%approach. We discuss the gauge invariance of the tight-binding under
%electromagnetic field and give a precise definition of g-factor.

\subsection{Tight-binding model of CdSe nanostructure}
We start from the empirical tight-binding model for the bulk CdSe
semiconductor with an $sp_3s^*$ basis.  The parameters we use for CdSe
are derived from the empirical parameters obtained by Lippens and
Lannoo \cite{lippens90} for the bulk CdSe in the zinc-blende
structure, assuming nearest-neighbor interactions only.  We construct
the CdSe nanocrystals with wurtzite structure corresponding to the
typical CdSe nanostructures seen in TEM images. \cite{shiang95} The
constructed structures have approximate but not exact $C_{3V}$
symmetry.  The same structures have been used in previous
time-independent tight-binding studies.
\cite{pokrant99,leung98,joshua03} We remove the dangling bonds on the
surface by shifting the energies of the corresponding hybrid orbitals
well above the conduction band edge. The spin-orbit interaction is
included in the Hamiltonian. Spin-orbit coupling constants are
assigned to both types of atoms, with $\lambda_{\mbox{Cd}}=0.151$eV
and $\lambda_{\mbox{Se}}=0.320$eV respectively. \cite{leung98,chadi77}
In order to reproduce the A-B splitting within the $sp_3s^*$ basis, a
crystal field of $-40$ meV is added to the $p_z$ local
orbitals. \cite{leung98}

\subsection{Time-dependent approach}
\label{timetb}
Instead of diagonalizing the tight-binding Hamiltonian directly we
employ the time-dependent approach which has been used previously to
calculate electronic and excitonic properties of CdSe nanocrystals
with zinc-blende structure. \cite{hill96} In the following we will
briefly describe the time-dependent tight-binding technique, indicate
the differences compared with previous calculations, and show its
advantage for calculation of g-factors.

The time-dependent method depends primarily on the spectral
decomposition for an arbitrary initial state. Let $|E_n\rangle$ be the
complete set of eigenfunctions of the Hamiltonian. Any initial state
$|\psi(0)\rangle$ can be expressed as the linear combination of
the eigenfunctions
\begin{equation}
|\psi(0)\rangle=\sum_n b_n |E_n\rangle .
\end{equation}
The wavefunction at a later time $t$ is
\begin{equation}
|\psi(t)\rangle=e^{-iHt}|\psi(0)\rangle=\sum_n b_n e^{-iE_nt}|E_n\rangle .
\end{equation}
Projecting the wavefunction at time $t$ onto the initial wavefunction
and perform the Fourier transform one finds
\begin{equation}
\int_{-\infty}^\infty dt e^{iEt} \langle \psi(0)|\psi(t)\rangle=
\sum_n |b_n|^2 \delta(E-E_n).
\end{equation}
Thus the resulting Fourier spectrum can give us the spectral weight of
the initial state in the eigenfunction basis and the eigenenergies of
the eigenstates, provided that the eigenstates have non-zero overlap
with the initial state.  To get the total density of states one can
sum over the spectral decompositions obtained using each wavefunction
in a complete set as the initial state in term. The natural and
convenient complete set to choose in the tight-binding framework is
the direct product set of all local site orbitals, atomic-orbitals,
and spin states. Then
\begin{equation}
\label{tdos_eq}
\sum_n \delta(E-E_n)= \sum_{il\sigma} \int_{-\infty}^\infty dt e^{iEt}
\langle \psi_{il\sigma}(0)|\psi_{il\sigma}(t)\rangle.
\end{equation}
where $|\psi_{il\sigma}(0)\rangle=|site,orbital,spin\rangle$.  To
achieve $\delta$-function resolution one would need to have the
infinite length record of the correlation function $\langle
\psi(0)|\psi(t)\rangle$. In practice only a finite length T of record
is available, which gives rise to artificial sidebands around a
broadened $\delta$-function approximation. The finite record length is
taken into account by multiplying the right hand size of
Eq.~(\ref{tdos_eq}) with the normalized Hamming window function
$w(t)$,\cite{feit82} where

\begin{equation}
\label{Hanning}
\begin{array}{crlll}
w(t) & = & 1-\cos(\frac{2\pi t}{T}), & \mbox{if} & 0 \leq t \leq T,\\  
     & = & 0,                        & \mbox{if} & t > T. \\ 
\end{array}
\end{equation}
The window function will reduce the sidelobes of the broadened
$\delta$-functions and generate a normalized peak height. The
resulting spectrum is of the form
\begin{equation}
\label{windowfunction}
\sum_n \mathcal{W}_n \mathcal{L}(E-E_n)= \sum_{il\sigma}
\int_{0}^\infty dt e^{iEt} w(t) \langle
\psi_{il\sigma}(0)|\psi_{il\sigma}(t)\rangle,
\end{equation}
where $\mathcal{W}_n$ represents the absolute spectral weight in
eigenstate $|E_n\rangle$ and the lineshape function
$\mathcal{L}(E-E_n)$ is defined by
\begin{equation}
\mathcal{L}(E-E_n)=
\frac{e^{i(E-E_n)T}-1}{i(E-E_n)T}-\frac{1}{2}\sum_{s=\pm 1}
\frac{e^{i(E-E_n)T+2 s \pi}-1}{i(E-E_n)T+2 s \pi}.
\end{equation}
If the total wavefunction propagation time is $T$, the energy
resolution is $\sim\Delta E=\pi/T$. If the energy difference between
the desired eigenenergy $E_n$, and adjacent eigenenergies is larger
than $\pi/T$, the spectrum near energy $E_n$ can be approximately
represented by $\mathcal{W}_n \mathcal{L}(E-E_n)$ with very high
accuracy. Assuming this form, the value of the eigenvalue can be
determined with accuracy much higher than $\pi/T$. To get the most
accurate value possible it is desirable to perform the time
integration of Eq.~(\ref{windowfunction}) by direct integration
instead of using a discrete Fourier transform.

%This would
%make an accurate identification of the eigenvalues possible and enalbe
%us to use the height of the peaks as an accurate way to counte the
%number of states.

In order to use the spectral method one must be able to calculate the
time propagator $e^{-iHt}$ efficiently. In order to accomplish this we
first break the time propagator into a series of short time
propagators $e^{-iHt}=(e^{-iHdt})^N$ with $t=N dt$. For the short time
propagator we make use of the Baker-Hausdorff formula\cite{sakurai}
expansion to obtain the expansion
\begin{eqnarray}
 e^{-iH dt}&=&e^{-i\left(H_1+\cdots +H_n\right)dt} \\ 
&\approx&
e^{-iH_1 dt}\cdots e^{-iH_n dt} e^{-iH_n dt}\cdots e^{-iH_n dt}+O(dt^3). \nonumber
\end{eqnarray}
To implement this decomposition we first break the tight-binding
Hamiltonian into the on-site self-energy terms, the local spin-orbit
terms, the local Zeeman terms, and the hopping terms. The on-site
spin-orbit interaction is diagonalized and exponentiated analytically
in the basis of the tight-binding orbitals, {\it i.e.} the $6 \times
6$ matrix of the p-orbitals with spin. For the hopping terms we
further use the checkerboard decomposition \cite{suzuki76,hill96} to
divide these to different independent directions. Note that in the
zinc-blende structure there are only four fundamental directions while
in the wurtzite structure there are seven fundamental directions.  As
a result of this decomposition each term contributing to the short
time propagator can consequently be evaluated analytically
\cite{hill96} and the time evolution of the state can be calculated
very efficiently.

The eigenfunction $|E_n\rangle$ with eigenenergy $E_n$ can be
calculated from
\begin{equation}
\label{eigenstate}
|E_n\rangle \propto \int_0^\infty dt e^{i E_n t}|\psi(t)\rangle,
\end{equation}
provided there is non-zero overlap between the initial wavefunction
and the desired eigenfunction, {\it i.e.} $\langle E_n|\psi(0)\rangle
\neq 0$. Typically when there is no magnetic field $(\mathbf{B}=0)$,
the initial state is taken as a uniform superposition of local
orbitals with specific angular momentum index. The resultant
eigenfunctions are then used as the starting point to calculate the
eigenfunctions and eigenenergies when the magnetic field is turned on.

If there are degenerate eigenstates, the right hand side of
Eq.~(\ref{eigenstate}) will in general be some unknown linear
combination of these eigenstates. However, if a set of exact or
approximate quantum numbers which can be used to label the degenerate
eigenstates are known in advance eigenfunctions corresponding to
definite quantum numbers can be derived by judiciously choosing an
initial state having the same quantum numbers. Typically the angular
momentum index is used in this work for this purpose. This property
will be used to generate Kramers' doublets in our calculations.

There are three important energy scales in this problem. The first
energy scale is the energy difference between lowest conduction
electron and higher energy conduction electrons and the difference
between highest valence hole and lower energy hole states. This energy
scale is typically at the order of 100 meV or larger. The second
energy scale is the energy difference between nearly degenerate hole
states that correspond approximately to the heavy hole and light hole
states in the bulk limit. This energy scale is size dependent and is
sensitive to the shape of the nanocrystals. In our calculation we find
this energy scale to be 1-100 meV. The last important energy scale is
the magnetic field induced splitting for a Kramers' doublet from which
the g-factors are extracted. Typically this energy scale ranges from
couple of hundred $\mu$eV to several $\mu$eV.

The maximal total propagation time is about 1280000 1/eV, resulting in
an energy resolution of 2.5 $\mu$eV. In our calculation this energy
resolution is enough to single out the spectrum of band edge electron
and hole states from other higher energy states. It is also sufficient
to resolve the two nearly degenerate hole states. Once a high
resolution eigenfunction is generated, by using window function
Eq.~(\ref{Hanning}-\ref{windowfunction}) to suppress the contribution
from adjacent eigenstates , the eigenenergies of band edge states can
be determined with accuracy up to 1 $\mu$eV.

%and to distinguish the (almost) degenerate heavy hole and light hole states.

\subsection{Calculation of the g-factors}
\label{g_def}
It is important to clarify the definition and the sign convention for
the g-factors, especially when these g-factors are anisotropic. When
there is no external magnetic field Kramers' theorem guarantees that
each eigenstate is at least two fold degenerate. In bulk CdSe the
heavy hole and light hole are degenerate at the $\Gamma$ point. In a
CdSe nanostructure it is expected that the quantum confinement will
lift this degeneracy. In this work the g-factors will be defined with
respect to the Kramers' doublet.  For a Kramers' doublet the effective
magnetic Hamiltonian has the form
\begin{equation}
\label{h_eff}
H_{\mbox{eff}}(\mathbf{B})=\mu_B \mathbf{B}\cdot \stackrel{\leftrightarrow}{\mathbf{G}}\cdot\tilde{\mathbf{S}},
\end{equation}
where $\tilde{\mathbf{S}}$ is the effective spin operator, which is
defined with respect to the two Kramer's states $|\psi^{\pm}\rangle$,
and $ \stackrel{\leftrightarrow}{\mathbf{G}}$ is the $3\times 3$
g-factor tensor.  In the Kramers' state basis, the effective spin
operator $\tilde{\mathbf{S}}$ has the form
\begin{equation}
\label{effective_s}
\tilde{\mathbf{S}}_x=\frac{m}{2}\sigma_x,
\tilde{\mathbf{S}}_y=\frac{m}{2}\sigma_y,
\tilde{\mathbf{S}}_z=\frac{m}{2}\sigma_z,
\end{equation}
where $m$ is an integer chosen so that the real and effective spin are
approximately equal. The $ \stackrel{\leftrightarrow}{\mathbf{G}}$
tensor can be diagonalized in the principal axes. Let $\hat{e}_i$,
$i=1,2,3$ be the principal axes. Then an external magnetic field
\begin{equation}
\mathbf{B}= B_1 \hat{e}_1+B_2 \hat{e}_2 +B_3 \hat{e}_3
\end{equation}
will give rise to a Zeeman splitting 
\begin{equation}
\Delta E(\mathbf{B})
\equiv \left(E(\mathbf{B})-E(0)\right)
=
\mu_B \sum_i g_i B_i .
\end{equation}
We will denote the values $g_i$ as principal g-factors. Individual
$g_i$ can be identified via varying external magnetic field along each
of the principal directions and calculating the field dependent Zeeman
splitting.

In our calculation for CdSe nanocrystal we find that the conduction
electron is primarily s-like. In this case one can identify the
effective spin operator with the real spin operator and give the
g-factors a definite sign. By choosing the initial wavefunction to
have a well defined real spin index, the resultant eigenfunction will
approximately have the corresponding effective spin index.

On the other hand, we find the hole to be primarily
p-like. Furthermore, there is a strong mixing between heavy hole and
light hole component which is sensitive to the size and the shape of
the nanostructures. It is thus difficulty to link the effective spin
to the real spin, resulting in an ambiguity of the sign of the hole
g-factors.  In this situation it is more appropriate to express the
Zeeman splitting by the quadratic form
\begin{equation}
\Delta E^2(\mathbf{B})
\equiv \left(E(\mathbf{B})-E(0)\right)^2
=
\mu_B^2 \sum_i g^2_i B^2_i,
\end{equation}
in which the sign of principal g-factor is not well defined. Typically
the sign convention of the g-factor in the atomic and bulk limit can
then be used as a convention to assign an definite sign to the
g-factors in the nanostructure. In both cases there usually exists a
simple relation between the effective spin operator and real spin
operator, which enables us to determine the corresponding sign of the
g-factor in the nanostructure. However there is a dearth for
experimental results of electron, hole, and exciton g-factor in bulk
CdSe. Hence it is difficult in this case to use bulk experimental
results as a guide to determine the sign of the g-factors.

In order to assign the sign to the hole g-factors for CdSe we
therefore adopt the following scheme. The hole wavefunction will be
calculated by propagating an initial state which has definite angular
momentum index $j=3,j_z=3/2, 1/2$.  The sign is then determined by
whether the magnetic field induced energy shift is positive or
negative. In the bulk limit this scheme will reproduce the heavy hole
and light hole value correctly.  In our calculation we find that the
electron g-factor decreases as the size increases. This represents
qualitatively the same trend as seen in effective mass type
calculations\cite{gupta02} for various semiconductor
nanostructures. On the other hand the calculated hole g-factor for
CdSe shows oscillations, making correlation with the atomic and bulk
limits more difficult.

\subsection{Gauge invariance}
Since in this work the g-factors will be determined via the energy
splitting of the electron and hole states under the external magnetic
field, it is critical to cast the tight-binding model into a gauge
invariant form. We use the Peierls-coupling tight-binding scheme to
ensure the gauge invariance in our tight-binding model.
\cite{graf95,boykin01_1,boykin01_2} In this scheme an electromagnetic
field specified by the scalar potential $\Phi(\vec{r},t)$ and the
vector potential $\mathbf{A}(\vec{r},t)$ will modify the on-site
$\langle \alpha, \mathbf{R}_i|H|\alpha, \mathbf{R}_i\rangle$ and
off-site $\langle \alpha^\prime, \mathbf{R}_i^\prime|H|\alpha,
\mathbf{R}_i\rangle$ tight-binding parameters via

\begin{equation}
\langle \alpha, \mathbf{R}_i|H|\alpha, \mathbf{R}_i\rangle \rightarrow
\langle \alpha, \mathbf{R}_i|H|\alpha, \mathbf{R}_i\rangle -\Phi(\mathbf{R}_i,t),
\end{equation}
and
\begin{equation}
\langle \alpha^\prime, \mathbf{R}_i^\prime|H|\alpha, \mathbf{R}_i\rangle \rightarrow
\langle \alpha^\prime, \mathbf{R}_i^\prime|H|\alpha, \mathbf{R}_i\rangle 
e^{-i\frac{e}{\hbar } \int_{R_i}^{R_i^\prime} \vec{A}(\vec{r},t)\cdot d\vec{l}} ,
\end{equation}
where a straight line should be taken for the vector potential
integral.  In this calculation, the on-site intra-atomic dipole matrix
elements $\langle \psi_{il\sigma} | \delta \vec{r}_i| \psi_{i l^\prime
\sigma^\prime} \rangle$ are set to zero. As pointed out by Foreman
\cite{foreman02}, this choice ensures gauge invariance but cannot
correctly describe intra-atomic transitions in the atomic
limit. However, predictive calculation of atomic transitions is not
contained within empirical tight-binding treatment. Previously,
optical absorption spectra have been dealt with in tight-binding
analysis by incorporating dipole matrix elements as extra fitting
parameters. For CdSe nanocrystals, calculation of the bulk absorption
spectrum has been found to be insensitive to the magnitude of the
on-site dipole matrix elements,\cite{leung97,leung98,pokrant99} This
provides some empirical justification for setting these matrix
elements to zero in order to achieve gauge invariance.

%However, this limit is not recovered in the generic empirical
%tight-binding treatment \cite{leung97} and the on-site dipole matrix
%elements can be treated as extra fitting parameters.  It is found in
%one of our previous calculation \cite{leung98} that setting all
%on-site dipole matrix elements zero except $\langle \psi_{Cd
%s^*}|\delta \vec{r}_i | \psi{Cd p_z}\rangle=0.4 \vec{e}_z$
%a.u. reproduce good agreement with bulk absorption spectrum while in
%another calculation \cite{pokrant99} that by setting all on-site
%dipole matrix elements to zero reproduce best agrrement with bulk
%absorption spectrum. Those results justify our choice of setting these
%matrix elements to zero.

For gauge-invariant correction we thus only need to modify the the
hopping constant between nearest neighbors. Since there are only
seven independent hopping directions in a wurtzite structure the gauge
phase can be calculated and stored before performing the time
propagation.  A brief summary of the gauge phase in the wurtzite
structure is given in the appendix \ref{appendix-1}.

To estimate the contribution to the Zeeman splitting from the gauge
phase, we have calculated the Zeeman splitting without gauge phase for
some of the nanocrystals. We find that the gauge phase contributes
10-40\% of the Zeeman energy. Without the gauge phase the Zeeman
splitting increases and becomes more isotropic.

%We find that the Zeeman splitting
%increases 15-45 \% and it become more isotropic.

\section{results}
\label{results}
We investigate CdSe nanostructures having 66-1501 atoms. This
roughly corresponds to the size range of 15-43 \AA\ in effective
diameter. In our calculation we define the aspect ratio to be the
ratio between effective in-plane diameter ($\sqrt{L_x L_y}$) and
out-of-plane diameter ($L_z$). The aspect ratio of these
nanostructures ranges from 0.68 to 1.64. These can be divided into
three different aspect ratio groups. The first group has aspect ratio
well below one, ranging from 0.68 to 0.85. The second group has aspect
ratio approximately equal to one, ranging from 0.99 to 1.09. The third
group has aspect ratio well above one, ranging from 1.37 to 1.64. A
nanostructure with aspect ratio 2.44 is also studied, in order to
probe the trends of g-factors in the quantum rod limit. In Table
\ref{size} we summarize the in-plane diameter and out-of-plane
diameter values of the nanostructures. If the aspect ratio of the
nanostructures deviate from 1.0 by less than 10\% then it is
appropriate to use a single effective diameter ($\sqrt{L_x L_y L_z}$)
to characterize the nanostructure. Note that the nanostructures used
in the TRFR experiment have reported aspect ratios in the range of
1.17-1.34. \cite{gupta02} However a single effective diameter was
nevertheless used to characterize the nanostructures. Furthermore, the
TRFR sample has 5-15\% size distribution and $\pm 0.2$ aspect ratio
variation.  Hence, one must be cautious when making a quantitative
comparison between the calculated and the experimental results.

\begin{widetext}

\begin{table}
\caption{\label{size} Size, diameters, and aspect ratio of the
nanostructures.}
\begin{tabular}{|c|c|c|c|c|c|c|c|c|c|c|c|c|} \hline
Number of atoms           
&66    &108   &144   &237   &336   &384   &450   &561   &758   &768   &777   &1501   \\ \hline
$\sqrt{L_x L_y}$ (\AA)    
&13.38 &13.38 &16.92 &21.85 &21.85 &25.39 &26.76 &22.85 &34.55 &27.42 &20.44 &43.01  \\ \hline
$L_z$  (\AA)              
&11.38 &18.38 &18.38 &14.88 &21.88 &24.38 &21.88 &35.88 &35.88 &39.38 &49.88 &42.88  \\ \hline
Aspect Ratio              
& 0.85 &1.37  &1.09  &0.68  &1.00  &0.99  &0.82  &1.64  &1.04  &1.44  &2.44  &0.99   \\ \hline
%$g_\parallel$             &1.9245&1.9004&1.8520&1.8347&1.8036&1.7483&1.7449&1.7310&1.6827&1.7000&1.7172&1.5998 \\ \hline
%$g_z$                     &1.6861&1.7202&1.6205&1.4650&1.4615&1.3752&1.3302&1.2853&1.0884&1.2524&1.2524&0.9503 \\ \hline
\end{tabular}
\end{table}

\end{widetext}

\begin{figure}
\includegraphics[scale=0.35,angle=270]{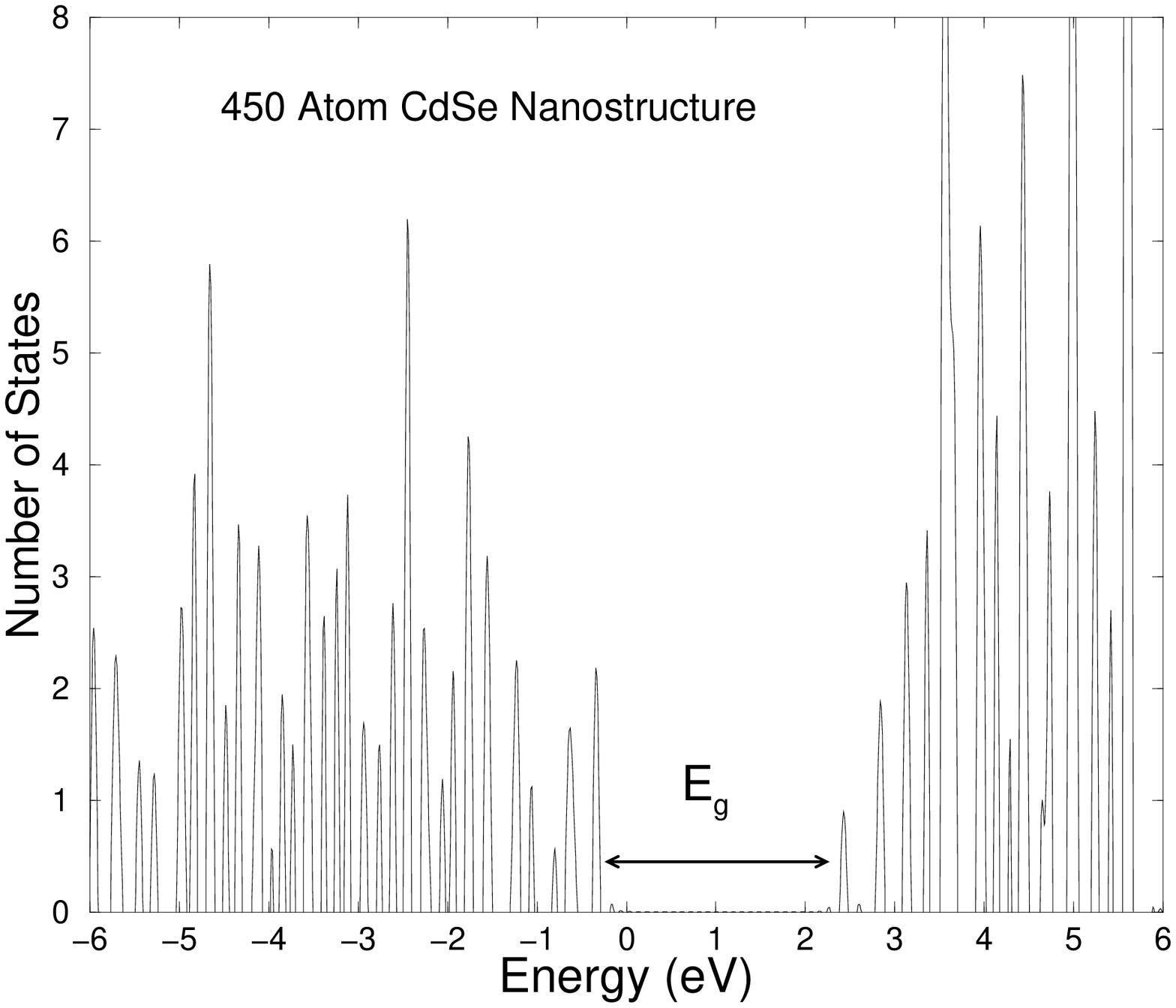}
\caption{\label{tdos}Total density of states for a 450 atom CdSe nanocrystal.}
\end{figure}

To verify that the tight-binding model can reproduce the general
features of conduction band, valence band, and identifiable band gap
for nanostructures we have calculated the total density of states
(TDOS) for smaller nanostructures (66-450 atoms). In Fig.~\ref{tdos}
we plot the low resolution ($\approx$ 50 meV) TDOS for a 450 atom CdSe
nanostructure. It is evident from the figure that the conduction band
edge (CBE), valence band edge (VBE), and band gap can be easily
identified. It should be noted that the TDOS calculation is
computationally expensive because one has to sum over a complete set
of initial states. However only the states at the band edges are
relevant to the optical orientation experiment. A prior knowledge of
the TDOS is not necessary for calculation of the band edge
eigenstates. A reasonable initial guess of the band edge eigenenergy
is sufficient for calculation of high resolution band edge
eigenenergies and eigenfunctions through an iterative procedure
described below. For the smaller nanostructures where we have
calculated the TDOS, we use the band edge energies identified from the
TDOS data as initial values. For the larger nanostructures, we assign
the initial value of band edge energies by extrapolating the band edge
energies of the smaller nanostructures.

\begin{figure}
\includegraphics[scale=0.30,angle=270]{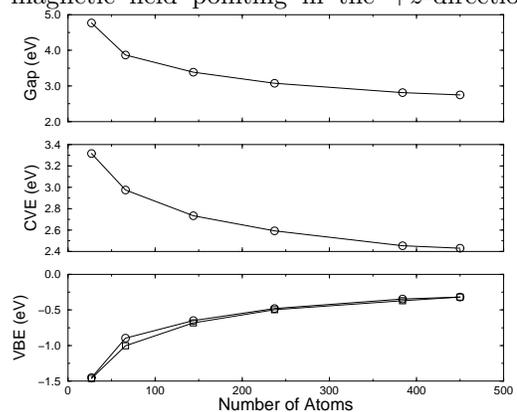}
\caption{\label{gap}(a) Band gap (b) CBE energy (c) VBE energy as a
function of the number of atoms. Note that the VBE consists of two
nearly degenerate levels in a CdSe nanocrystal, each corresponding to
a perturbed Kramers' doublet, {\it i.e.} 4 states in total.}
\end{figure}

To get the high resolution band edge eigenenergies and eigenstates we
first estimate the eigenenergies as described above.  A low resolution
eigenstate is then generated using some judiciously chosen initial
state. The initial state is set up to have non-zero overlap with the
desired eigenfunction and to posses a well-defined value of some
particular quantum number such as the z-component of the local total
angular momentum $|j,j_z\rangle$.  This low resolution eigenstate is
then put through the spectral weight analysis described in
section \ref{timetb} which results in a higher resolution
eigenenergy. The higher resolution eigenenergy is then used together
with the lower resolution eigenstate to generate a higher resolution
eigenstate. This process is iterated until the desired accuracy is
acquired and, in the case of the hole, until the near degeneracy
between heavy hole like and light hole like doublets is lifted. Once
the CBE and VBE eigenenergies are found, the band gap can be trivially
calculated, from $E_{\mbox{Gap}}=E_{\mbox{CBE}}-E_{\mbox{VBE}}$. In
Fig.~\ref{gap} we plot the high resolution results for the size
dependent CBE energy, VBE energy, and band gap. These results are all
stable with respect to further iteration. Note that the VBE consists
of two nearly degenerate Kramers' doublets. As the size of the
nanostructure increases, these two doublets will converge respectively
to the heavy and light hole doublets in bulk CdSe.

%In order to enable us to make a reasonable connection to the
%condunction electron, heavy hole, and light hole in the bulk CdSe
%semiconductor we assign a total angular moment index to the
%eigenfunctions calculated in the nanostructure. For each eigenfunction
%it's spectra weight in terms of local orbitals are calculated. The
%eigenfunction will be identify with the majority local orbital with
%definite angular momentum index.

\subsection{Electron g-factor}

Although the CdSe nanostructures we studied here have only approximate
but not exact $C_{3v}$ symmetry, we expect nevertheless that the
principal axes are still located approximately along the $x$, $y$, and
$z$ directions. This is supported by the result of a perturbative
time-independent tight-binding calculation of g-factors for these same
nanocrystals \cite{joshua03} in which it was found that $g_x \approx
g_y\neq g_z$. To accurately identify the Zeeman splitting it is
necessary to generate the Kramers' doublet which will evolve into the
Zeeman eigenstates when we turn on the external magnetic field. (This
is essentially equivalent to solving the zeroth order degenerate
perturbation problem.) The Kramers' doublet $|\psi^{\pm z} \rangle$
for a magnetic field pointing in the +$z$-direction can be generated
via setting all local orbitals of the initial states to have spin
equal to $\pm \frac{1}{2}$. The Kramers' doublet for x and y
directions are then calculated as $|\psi^{\pm
x}\rangle=\frac{1}{\sqrt{2}}(|\psi^{+z}\rangle \pm |\psi^{-z}\rangle)$
and $|\psi^{\pm y}\rangle=\frac{1}{\sqrt{2}}(|\psi^{+z}\rangle \pm i
|\psi^{-z}\rangle)$ respectively. The external magnetic field is
limited to be less than 10 Tesla, which corresponds to the range of
magnetic field in the typical experiments \cite{gupta02,gupta99}. To
make the connection to the CBE in the bulk material, which is s-like,
we calculate the spectral weight of the $|\psi^{+z}\rangle $ state in
the $|s,\sigma=\frac{1}{2}\rangle$ local orbital of Cd and Se. In
Table \ref{spectral_s} we summarize the size dependence of these
s-orbital spectral weights.  We find that the CBE electron in the
nanostructure is still primarily s-like, with spectral weights greater
than 0.75 for all sizes. The s-orbital contribution increases
monotonically as the size increases.

\begin{widetext}

\begin{table}
\caption{\label{spectral_s} Spectral weight of $|\psi^{+ z}\rangle$ in
the $|s,\sigma=\frac{1}{2}\rangle$ local orbital of Cd and Se.}
\begin{tabular}{|c|c|c|c|c|c|c|c|c|c|c|c|c|} \hline
Number of atoms                & 66   & 108  & 143  & 237  & 336  & 384  & 450  & 561  & 758  & 768  & 777  & 1501 \\ \hline
$|s,\sigma=\frac{1}{2}\rangle$ & 0.75 & 0.79 & 0.81 & 0.83 & 0.85 & 0.86 & 0.86 & 0.88 & 0.87 & 0.88 & 0.89 & 0.90 \\ \hline
\end{tabular}
\end{table}

\end{widetext}

\begin{figure}
\includegraphics[scale=0.35,angle=270]{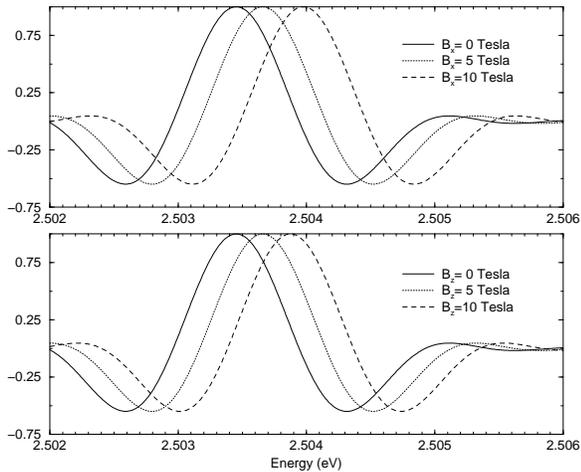}
\caption{\label{zeeman}Zeeman shift of the $|\Psi^{+x}\rangle$
component of the Kramers' doublet in the CBE when (a) the external
field B is in the +$x$-direction and (b) the +$z$-direction.}
\end{figure}

\begin{figure}
\includegraphics[scale=0.35,angle=270]{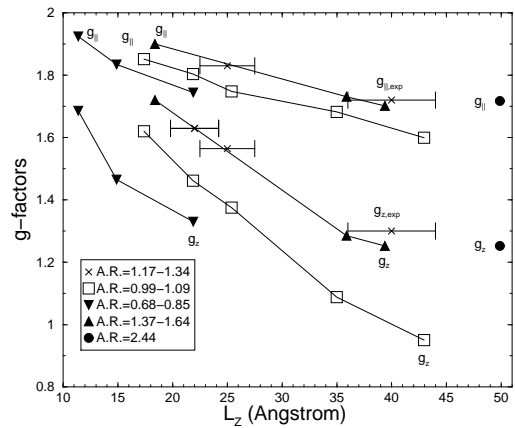}
\caption{\label{e_gfactor}CdSe nanocrystal electron g-factors as a
function of the nanocrystal length parameter $L_z$.}
\end{figure}

In Fig.~\ref{zeeman} we plot the magnetic field dependent spectra for
the $|\psi^{+x}\rangle$ state with magnetic field in the $x$-direction
and for the $|\psi^{+z}\rangle$ state with magnetic field in the
$z$-direction, for a 336 atom CdSe nanocrystal. Assuming the spectral
peaks have lineshapes of the form $\mathcal{L}(E-E_n(B))$, the
magnetic field dependent eigenenergy $E_n(B)$ can then be determined
with very high accuracy. The g-factor is then extracted by fitting
$E_n(B)$ as a function of B. In Fig.~\ref{e_gfactor} we plot the
resulting electron g-factors as a function of the length parameter
$L_z$. The data are grouped according to the aspect ratio of the
nanostructure. Group 1 (down triangles) has aspect ratio 0.68-0.85,
group 2 (open squares)has aspect ratio 0.99-1.09, and group 3 (up
triangles) has aspect ratio 1.37-1.64. One calculation for nanocrystal
with aspect ratio 2.44 is also included (closed circle). The extracted
g-factors from TRFR experiments \cite{gupta02} are also plotted for
comparison (asterisks). Note that the size distribution of the sample
in the TRFR experiments is about 5-15\%, which is represented in the
figure by the horizontal error bar. The aspect ratio of the sample in
TRFR experiment in this size range is about 1.17-1.34, with a $\pm
0.2$ variation.

From the calculations we find that $g_x=g_y >g_z$ for all the
nanostructures. As a result only two sets of data are shown in the
figure and $g_\parallel$ is used to represent both $g_x$ and
$g_y$. The results show strong anisotropy between $g_z$ and
$g_\parallel$.  Both g-factors decrease monotonically as a function of
the size of the nanostructure. The value of $g_z$ decreases rapidly,
while the value of $g_\parallel$ decreases more
gradually. Fig.~\ref{e_gfactor} shows that our results are in good
agreement with the experimental values. It is evident from the figure
that when there are two distinct g-factors observed for a given
nanostructure, they should be identified with the in-plane
($g_\parallel$) and out-of-plane ($g_z$) g-factors of the
electron. This assignment is very different from the original
experimental suggestion that one of the g-factors should be identified
with the isotropic electron g-factor while the other g-factor might be
identified with an exciton g-factor.\cite{gupta02}

It is intriguing to look into the aspect ratio dependence of the
g-factors in more detail. We observe from Fig.~\ref{e_gfactor} that
both $g_\parallel$ and $g_z$ increase as the aspect ratio increases,
provided that the aspect ratio is less than 1.64.  We find that $g_z$
is more sensitive to the aspect ratio and increases much more with
this than $g_\parallel$. The g-factors begin to saturate between
aspect ratio 1.64 and 2.44. It is expected that if one continues to
increase the aspect ratio then $g_z$ should begin to decrease, since
it eventually should approach the bulk value. On the other hand
$g_\parallel$ should stay roughly constant after it saturates,
provided that the in-plane cross-section is kept the same when one
increases the aspect ratio. It should be emphasized that the aspect
ratio is only a simple indicator for the shape of the
nanostructure. Two nanostructures with similar number of atoms and
aspect ratio values might still have very different shape or surface
structure. From the observations above it is clear that we can
identify a range of aspect ratios in which the effect of anisotropy of
the wurtzite structure and that of the shape of the nanocrystal
partially cancel each other so that the electron g-factors become more
isotropic. In the cases studied here it appears that the cancellation
is not complete. It also appears unlikely from these exact calculation
of g-factors that the cancellation will become more complete for
large-size nanocrystals since the difference between $g_z$ and
$g_\parallel$ increases for larger nanocrystals having aspect ratio
approximately unity. As a result a true quasi-spherical regime as
predicted by an analysis perturbative in
spin\cite{rodina02_2,joshua03} in which the electron g-factors become
isotropic may never reached. Under certain growth conditions it is
possible to synthesis CdSe nanostructures with zincblende
structure.\cite{manna02} It is expected that for zincblende CdSe
nanocrystals, if the shape of the nanostructure has high symmetry then
there will be only one isotropic g-factor.\cite{lee}

\subsection{Hole g-factor}

We have calculated the values of $g_z$ for the two nearly degenerate
valence band edge doublets, which will be denoted by $h_1$ and
$h_2$. We define $h_1(h_2)$ to be the highest (second highest) energy
valence band state. To connect the $h_1$ and $h_2$ states to the heavy
and light hole states in the bulk material we calculate their spectral
weight in local orbitals possessing definite angular momentum quantum
number $|j=\frac{3}{2},j_z\rangle$.  In Table-\ref{spectral_h1}
(Table-\ref{spectral_h2}) we summarize the spectral weight of the
$h_1(h_2)$ states. We find that for the nanocrystals in the size range
we are interested in, the mixing between the $\pm \frac{3}{2}$ and
$\pm \frac{1}{2}$ components is very strong. The mixing also seems to
be sensitive to the size of the nanocrystal, without any clear trend
emerging.  As a result it becomes improper to rigorously identify the
$h_1(h_2)$ state with the heavy(light) hole states respectively, and
we therefore set $m=1$ in Eq.~(\ref{effective_s}) for all hole states.

\begin{widetext}

\begin{table}
\caption{\label{spectral_h1}Spectral weight of $h_1$ state in the
local orbitals $|j=\frac{3}{2},j_z\rangle.$ (See text.)}
\begin{tabular}{|c|c|c|c|c|c|c|c|c|c|c|c|c|c|} \hline
Number of Atoms    & 66   & 108  & 144  & 237  & 336  & 384  & 450  & 561  & 758  & 768  & 777  & 1501 \\ \hline
$j_z=+\frac{3}{2}$ & 0.47 & 0.03 & 0.27 & 0.82 & 0.06 & 0.30 & 0.81 & 0.06 & 0.87 & 0.37 & 0.51 & 0.88 \\ \hline
$j_z=+\frac{1}{2}$ & 0.12 & 0.78 & 0.24 & 0.02 & 0.78 & 0.21 & 0.03 & 0.79 & 0.02 & 0.11 & 0.34 & 0.03 \\ \hline
$j_z=-\frac{1}{2}$ & 0.29 & 0.00 & 0.37 & 0.04 & 0.04 & 0.39 & 0.06 & 0.05 & 0.04 & 0.18 & 0.02 & 0.04 \\ \hline
$j_z=-\frac{3}{2}$ & 0.01 & 0.03 & 0.02 & 0.00 & 0.04 & 0.02 & 0.02 & 0.02 & 0.00 & 0.25 & 0.00 & 0.00 \\ \hline
\end{tabular}
\end{table}

\begin{table}
\caption{\label{spectral_h2}Spectral weight of $h_2$ state in the
local orbitals $|j=\frac{3}{2},j_z\rangle.$ (See text.)}
\begin{tabular}{|c|c|c|c|c|c|c|c|c|c|c|c|c|c|} \hline
Number of Atoms    & 66   & 108  & 144  & 237  & 336  & 384  & 450  & 561  & 758  & 768  & 777  & 1501 \\ \hline
$j_z=+\frac{3}{2}$ & 0.02 & 0.46 & 0.04 & 0.46 & 0.64 & 0.40 & 0.36 & 0.53 & 0.37 & 0.37 & 0.35 & 0.04 \\ \hline
$j_z=+\frac{1}{2}$ & 0.06 & 0.20 & 0.03 & 0.03 & 0.10 & 0.13 & 0.13 & 0.33 & 0.17 & 0.11 & 0.10 & 0.83 \\ \hline
$j_z=-\frac{1}{2}$ & 0.08 & 0.13 & 0.37 & 0.35 & 0.09 & 0.17 & 0.41 & 0.02 & 0.37 & 0.19 & 0.43 & 0.02 \\ \hline
$j_z=-\frac{3}{2}$ & 0.68 & 0.07 & 0.43 & 0.00 & 0.08 & 0.22 & 0.01 & 0.01 & 0.01 & 0.23 & 0.06 & 0.06 \\ \hline
\end{tabular}
\end{table}

\end{widetext}

\begin{figure}
\includegraphics[scale=0.36,angle=0]{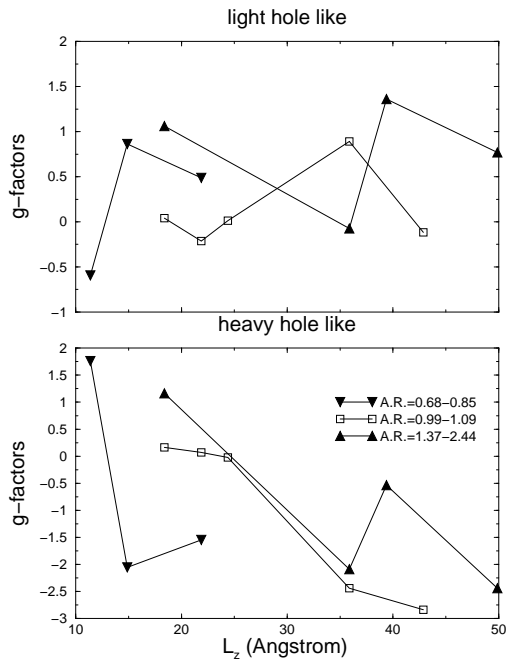}
\caption{\label{h_gfactor}CdSe nanocrystal hole g-factors as a
function of the nanocrystal length parameter $L_z$.}
\end{figure}

In Fig.\ref{h_gfactor} we plot the size dependent hole g-factors for
those two hole doublets. The data has been regrouped into heavy-hole
like and light-hole light states. For each size of the nanostructure
we look at the spectral weight of $h_1$ and $h_2$ states and determine
which is more heavy-hole-like. Similarly to the electron g-factors,
the data for hole g-factors are also grouped by the nanocrystal aspect
ratio. We observe that the two hole states have very different
g-factors. Both holes show strong oscillations as a function of the
size. It is important to clarify how these g-factors should approach
the relevant bulk values when the size of the nanocrystal
increases. In the bulk CdSe semiconductor, the valence band near the
$\Gamma$ point can be described by the Luttinger
Hamiltonian.\cite{luttinger55} The states $|\frac{3}{2},\pm
\frac{3}{2}\rangle $ and $|\frac{3}{2},\pm \frac{1}{2}\rangle $ are
associated with the heavy-hole and light-hole respectively. If the
heavy-hole and light-hole are really degenerate, then
Eq.~(\ref{h_eff}) is actually not appropriate, since the nature of the
$J=\frac{3}{2}$ angular momentum has to be taken into account. If the
heavy-hole and light-hole are not degenerate and the mixing between
$|\frac{3}{2},\pm \frac{1}{2}\rangle $ and $|\frac{3}{2},\pm
\frac{3}{2}\rangle $ components are small, then one can use
Eq.~(\ref{h_eff}) and set $m=3$ for the heavy-hole and $m=1$ for the
light-hole in Eq.~(\ref{effective_s}).

We speculate that the irregular mixing of local orbitals evident in
Tables \ref{spectral_h1} and \ref{spectral_h2} is the cause of the
hole g-factor oscillation in nanocrystal size. From the table we
observe that the $h_1$ state becomes more and more heavy-hole like for
nanocrystal with more than 450 atoms, provided that the aspect ratio
is close to one. It is expected that as the size of the nanocrystal
increases, the two hole states will eventually converge to the
heavy-hole and light-hole respectively. One must then be careful when
comparing the hole g-factors values calculated here with the bulk
heavy-hole g-factor since the latter is usually defined with $m=3$ in
Eq.~(\ref{effective_s}).\cite{nenashev03}

\section{summary and discussion}
\label{sum}

We have calculated the g-factors of the conduction band edge electrons
and valence band edge holes in CdSe nanostructures using the
time-dependent tight-binding method. This allows an exact,
non-perturbative analysis, with extremely high resolution of the
Zeeman shifts. We observe that the electron g-factors are strongly
anisotropic, with $g_x=g_y >g_z$ for all nanocrystal sizes. The size
dependence of the anisotropic electron g-factors agree quantitatively
with the values extracted from TRFR experimental data. This leads to
the conclusion that when there are two distinct g-factors in the TRFR
experiments, they should be identified with the electron in-plane
($g_\parallel$)and out-of-plane ($g_z$)g-factors respectively. This is
very different from the original experimental suggestion that one of
the g-factors is an isotropic electron g-factor while the other one is
an exciton g-factor.\cite{gupta02} We have investigated the aspect
ratio dependence of the electron g-factors.  We find that in general
the g-factors initially increase as a function of the aspect ratio,
and that $g_z$ increases more than $g_\parallel$.  The increase of
g-factors saturates around aspect ratio 1.62-2.44. It is expected that
$g_z$ will then begin to decrease again until it reaches the bulk
value while $g_\parallel$ will stay roughly the same, if the aspect
ratio continue to increase. The aspect ratio dependence allows us to
identify a regime where the anisotropy derived from the wurtzite
structure and that derived from the shape of the nanocrystal cancel
each other partially, resulting a more isotropic regime. However a
full cancellation is never reached in our calculation, unlike the
previous observation from a perturbative analysis. \cite{joshua03} It
also appears unlikely that in larger nanocrystals the cancellation
will become complete, since the difference between $g_z$ and
$g_\parallel$ for an unit aspect ratio nanocrystal increases as a
function of the size.

We find that the valence band edge consists of two nearly degenerate
Kramers' doublets, {\it i.e.} only a small perturbation from the bulk
states. The hole g-factors for these states show oscillations as a
function of the size. We speculate this is due to the strong size
sensitive heavy/light hole mixing of the two hole states.

One possible extension of the current calculation scheme is the
evaluation of the exciton g-factors in these nanostructures. The
g-factor of a uncorrelated electron-hole pair can be approximated by
$g_x=g_e-g_h$. However the Coulomb interaction gives rise to excitonic
effects and this simple picture then breaks down. The time-dependent
tight-binding method has been successfully applied to calculate
excitonic properties in CdSe nanocrystal \cite{hill96} and it appears
feasible to extend the current scheme to now calculate the
corresponding exciton g-factors. Since the exciton fine structure
splitting in CdSe nanocrystal is of the order 1-10 meV \cite{leung98}
the energy resolution obtained here ($\approx$ 1 $\mu$V)is enough to
resolve the exciton fine structure.

\begin{acknowledgements}
%\section{acknowledgements}
The authors acknowledge the financial support by the DARPA and ONR
under Grant No. FDN0014-01-1-0826 and by the ARO under Grant
No. FDDAAD19-01-1-0612. We thank NPACI for a generous allocation of
supercomputer time at the San Diego Supercomputer Center. We also
thank Dr. Seungwon Lee and Josh Schrier for fruitful discussions.
\end{acknowledgements}

\appendix*
\section{}
\label{appendix-1}
In the wurtzite structre there are 7 independent electron hopping
directions.  In this calculation these 7 hopping directions are
denoted by $\vec{d}_{dir}$, $dir=1 \cdots 7$, and are assigned to be
the following vectors:
\begin{equation}
\begin{array}{ll}
\vec{d}_1=\frac{a_0}{3}(0,0,3), & \\
\vec{d}_2=\frac{a_0}{3}(2\sqrt{2},0,-1),         & \vec{d}_5=\frac{a_0}{3}(-2\sqrt{2},0,-1), \\
\vec{d}_3=\frac{a_0}{3}(-\sqrt{2},\sqrt{6},-1),  & \vec{d}_6=\frac{a_0}{3}(\sqrt{2},-\sqrt{6},-1), \\
\vec{d}_4=\frac{a_0}{3}(-\sqrt{2},-\sqrt{6},-1), & \vec{d}_7=\frac{a_0}{3}(\sqrt{2},\sqrt{6},-1).
\end{array}
\end{equation}
Here $a_0=2.625$\AA\ is the lattice constant.  This convention enables
us to calculate the gauge-dependent quantities explicitly.

For a fixed external magnetic field $\vec{B}=(B_x,B_y,B_z)$ we assign
the vector potential to be $\vec{A}=\frac{1}{2} \vec{B}\times
\vec{r}$. 
%Since most of the time we only turn on the magnetic in one
%of the directions we also defined $\vec{A}_x=\frac{1}{2} (B_x,0,0)
%\times \vec{r}$ and similarly $\vec{A}_y$ and $\vec{A}_z$.  
We define a magnetic-field-dependent gauge phase
\begin{equation}
\phi(\vec{R}_a, \vec{d}_{dir})=
\frac{e}{\hbar} \int_{\vec{R}_a}^{\vec{R}_a+\vec{d}_{dir}}
\vec{A} \cdot \vec{dl}, 
\end{equation}
where $\vec{R}_a$ represents the position vector of an anion.  The
gauge phase for the cation can be easily calculated by taking the
appropriate complex conjugation of the phase of the corresponding
anion. With the line of integration to be a straight line, the
integral can be calculated analytically. Using this notation the
gauge-dependent short-time propagation of the electron hopping in some
particular direction becomes
\begin{eqnarray}
& & e^{-\hat{V}_{lm,dir} dt}
\left( A_l |\Psi_l (\vec{R}_a)\rangle + A_m |\Psi_m (\vec{R}_c)\rangle \right) \\
&=& 
\left( A_l \cos(V_{lm} dt)- iA_m\sin(V_{lm}dt)e^{+i\phi(\vec{R}_a,\vec{d}_{dir})} \right) |\Psi_l (\vec{R}_a)\rangle \nonumber \\
&+& 
\left( A_m \cos(V_{lm} dt)- iA_l\sin(V_{lm}dt)e^{-i\phi(\vec{R}_a,\vec{d}_{dir})} \right) |\Psi_m (\vec{R}_c)\rangle, \nonumber
\end{eqnarray}
where $V_{lm}$ is the hopping constant in zero magnetic field.

\end{document}